\documentclass[english,aps,prl,twocolumn,showpacs,superscriptaddress,groupedaddress,footinbib,reprint,noshowpacs]{revtex4}
\usepackage{graphicx}
\usepackage{amsmath}
\usepackage{comment}
\usepackage{amssymb}

\usepackage{subfigure}

\usepackage{subfigure}
\usepackage{color}
\usepackage{soul}

\begin{document}

\title{Scaling analysis of the screening length in concentrated electrolytes} 

\author{Alpha A. Lee}
\email{alphalee@g.harvard.edu}
\affiliation{John A. Paulson School of Engineering and Applied Sciences, Harvard University, Cambridge, MA 02138}

\author{Carla Perez-Martinez}
\affiliation{Department of Chemistry, Physical and Theoretical Chemistry Laboratory, University of Oxford, Oxford OX1 3QZ, U.K.}

\author{Alexander M. Smith}
\affiliation{Department of Chemistry, Physical and Theoretical Chemistry Laboratory, University of Oxford, Oxford OX1 3QZ, U.K.}
\affiliation{Department of Inorganic and Analytical Chemistry, University of Geneva, 1205 Geneva, Switzerland} 

\author{Susan Perkin}
\email{susan.perkin@chem.ox.ac.uk} 
\affiliation{Department of Chemistry, Physical and Theoretical Chemistry Laboratory, University of Oxford, Oxford OX1 3QZ, U.K.}

\begin{abstract}
The interaction between charged objects in an electrolyte solution is a fundamental question in soft matter physics. It is well-known that the electrostatic contribution to the interaction energy decays exponentially with object separation. Recent measurements reveal that, contrary to the conventional wisdom given by classic Poisson-Boltzmann theory, the decay length increases with ion concentration for concentrated electrolytes and can be an order of magnitude larger than the ion diameter in ionic liquids. We derive a simple scaling theory that explains this anomalous dependence of the decay length on ion concentration. Our theory successfully collapses the decay lengths of a wide class of salts onto a single curve. A novel prediction of our theory is that the decay length increases linearly with the Bjerrum length, which we experimentally verify by surface force measurements. Moreover, we quantitatively relate the measured decay length to classic measurements of the activity coefficient in concentrated electrolytes, thus showing that the measured decay length is indeed a bulk property of the concentrated electrolyte as well as contributing a mechanistic insight into empirical activity coefficients.
%It is well-known that the electrostatic contribution to the interaction energy decays exponentially with object separation. Recent measurements reveal that, contrary to the conventional wisdom given by classic Poisson-Boltzmann theory, the decay length increases with ion concentration for concentrated electrolytes and can be an order of magnitude larger than the ion diameter in ionic liquids. We derive a simple scaling theory for the decay length that explains this anomalous dependence of the screening length on ion concentration. Our theory successfully collapses the screening lengths of a wide class of salts onto a single curve. A novel prediction of our theory is that the screening length increases linearly with the Bjerrum length, which we experimentally verify by surface force measurements. Moreover, we quantitatively relate the measured screening length to classic measurements of the activity coefficient in concentrated electrolytes, thus showing that the measured screening length is indeed a bulk property of the concentrated electrolyte as well as contributing a mechanistic insight into empirical activity coefficients.

\end{abstract}

\makeatother
\maketitle

The structure of electrolytes near a charged surface and the resulting force between charged surfaces in an electrolyte solution is a fundamental question in soft matter physics. This question also underpins a plethora of applications, from supercapacitors \cite{conway2013electrochemical} to colloidal self-assembly \cite{evans1999colloidal}. The classic Debye-H\"{u}ckel theory \cite{huckel1923theory}, valid only for dilute electrolytes, predicts that the interaction between two charged surfaces in an electrolyte decays exponentially with the surface separation \cite{israelachvili2011} with a decay length, called the Debye length, given by
\begin{equation} 
\lambda_D =\sqrt{\frac{\epsilon k_B T}{4 \pi q^2 c_\mathrm{ion}}} \equiv \frac{1}{ \sqrt{4 \pi l_B  c_\mathrm{ion} }}, 
\label{debyehuckel}
\end{equation}
where $\epsilon$ is the dielectric constant of the medium (which is ion concentration-dependent), $k_B$ the Boltzmann constant, $T$ the temperature, $q$ the ion charge, $ c_\mathrm{ion}$ the ion concentration, and 
\begin{equation}
l_B = \frac{q^2 }{\epsilon k_B T}
\end{equation}
is the Bjerrum length. The Bjerrum length is the distance at which the interaction energy between two ions in a dielectric medium with dielectric constant $\epsilon$ equals the thermal energy unit $k_B T$. The Debye-H\"{u}ckel theory is a mean-field theory for asymptotically dilute electrolytes, i.e. where $l_B^3  c_\mathrm{ion} \ll 1$, so that the ion-ion separation is far greater than the Bjerrum length and thus the Coulomb interactions can be treated as a perturbation to ideal gas behaviour. 

The physical picture is less clear for concentrated electrolytes: Recent surface force balance (SFB) studies show that the interaction force between charged surfaces in an ionic liquid (molten salt at room temperature) decays exponentially, but with a decay length that is orders of magnitude longer than the Debye length or the ion diameter \cite{gebbie2013ionic, espinosa2013microslips, gebbie2015long, smith2016electrostatic}. The screening lengths in concentrated inorganic salts are also long and increase with electrolyte concentration \cite{smith2016electrostatic}, in direct opposition to the prediction of the Debye-H\"{u}ckel theory. Therefore, the anomalously long screening length is not a curiosity associated with ionic liquid chemistry, but appears to be an universal feature of Coulomb interactions between ions. Rigorous perturbative extensions of mean-field Poisson-Boltzmann equation have been developed (see refs \cite{attard1996electrolytes,nonner2000binding,levin2002electrostatic,naji2013perspective} for comprehensive reviews). However, although those theories predict an increase in screening length for increasing ion concentration, the magnitude of the screening length and its dependence on ion concentration cannot be explained by those pioneering works. 

In this Letter, we will first motivate the scaling theory by showing that the screening length measurements reported in the literature for diverse classes of electrolytes can be collapsed onto a single curve, and verify the scaling relationship with a new set of SFB measurements where the Bjerrum length is varied by changing the solvent but fixing other variables. We will then relate the measured screening length with classic measurements of the activity coefficient as a function of ion concentration to show that the screening length is a property of the bulk electrolyte. Finally, we will derive a simple scaling theory that explains the dependence of the screening length on ion concentration and Bjerrum length. 

\begin{figure}
\centering
\includegraphics[scale=0.8]{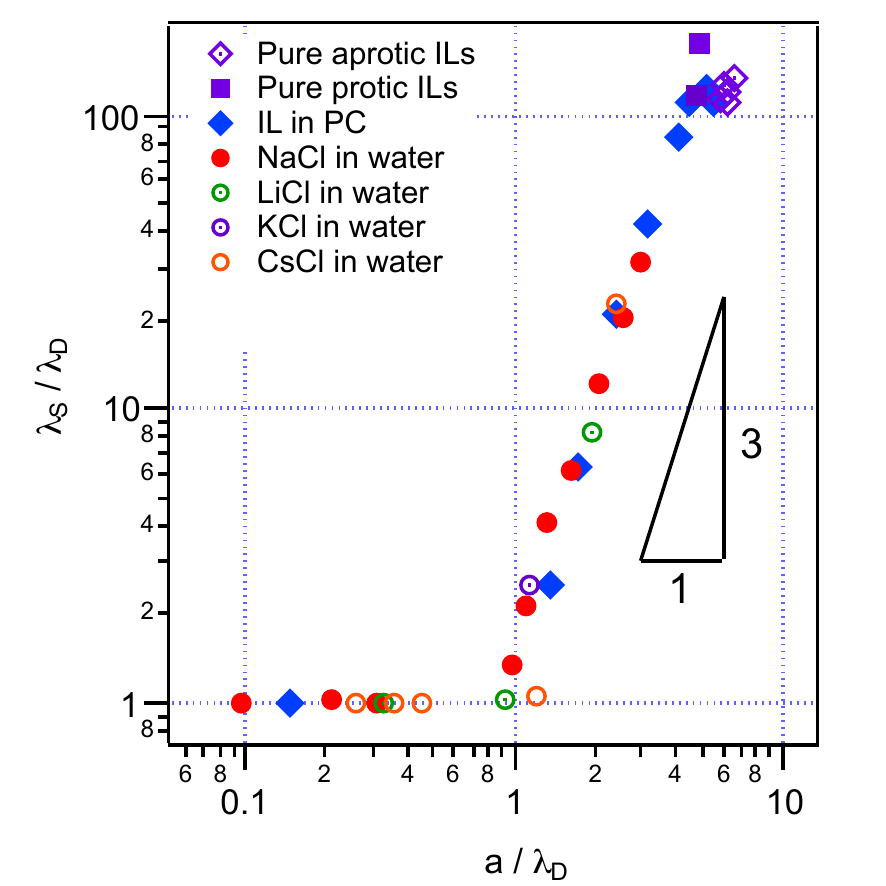} 
\caption{Experimentally measured screening length, $\lambda_S$ \cite{israelachvili1984measurement,baimpos2014effect,smith2016electrostatic,hjalmarsson2017switchable,lee2017FD}, normalised by $\lambda_D$, plotted against $a/\lambda_D$ for a range of pure ionic liquids (ILs), ionic liquid mixed with propylene carbonate (PC) molecular solvent, and various 1:1 inorganic salts in water. One point for a pure protic ionic liquid, ethylammonium nitrate, arises from a new surface force balance measurement by us. See Supplemental Material for a linear-linear plot.}
\label{Fig:ExperimentalFigureThree}
\end{figure}

We begin with a simple exercise in dimensional analysis: there are four salient lengthscales in the system, the experimentally measured screening length $\lambda_S$, $\lambda_D$, $l_B$, and the ion diameter $a$. We emphasize that the experimentally measured screening length $\lambda_S$ is extracted by fitting the interaction force between charged surfaces as a function of surface separation to an exponential decay; the fact that $\lambda_S$ can be orders of magnitude larger than $\lambda_D$ is a key result of this Letter.  Motivated by the seminal works by Kirkwood \cite{kirkwood1934theory,kirkwood1936statistical} and subsequent molecular dynamics simulations \cite{keblinski2000molecular}, a reasonable choice of dimensionless quantities is $\lambda_S/\lambda_D$ and $a/\lambda_D$ as the Bjerrum length is already in the Debye length. Figure \ref{Fig:ExperimentalFigureThree} shows that the screening lengths, $\lambda_S$, of different electrolyte solutions agree with the Debye-H\"{u}ckel result (Equation (\ref{debyehuckel})) when $a/\lambda_D \ll1$, but follows a universal scaling relationship
\begin{equation}
\frac{\lambda_S}{\lambda_D} \sim \left( \frac{a}{\lambda_D}\right)^3 
\label{scalinglaw}
\end{equation}
in the concentrated regime $a/\lambda_D > 1$. The dielectric constant of alkali halide solutions are taken from dielectric spectroscopy studies \cite{hasted1948dielectric,wei1992ion,buchner1999dielectric}, and the dielectric constants of ionic liquid-solvent mixtures are estimated using effective medium theory \cite{Bergman1978}. For ion size estimates, we take the mean  ion diameter of alkali halides \cite{shannon1976revised}. The complex geometry of ionic liquid ions precludes a clear-cut definition of ion diameter, and we estimate the ion diameter as $c_{\mathrm{pair}}^{-1/3}/2$ of the pure ionic liquid, where $c_{\mathrm{pair}} = c_{\mathrm{ion}}/2$ is the ion pair concentration. We note that the salts considered here are univalent and thus $q=1$.

We note that the data in Figure \ref{Fig:ExperimentalFigureThree} covers a wide range of ion sizes and chemical functional groups, thus the scaling relationship is a result of electrostatic interactions rather than specific ion chemistry. The physical significance of Equation (\ref{scalinglaw}) is perhaps more transparent when rearranged as 
\begin{equation}
\lambda_S \sim l_B a^3 c_{\mathrm{ion}}. 
\label{scalinglaw_dim}
\end{equation}
The dependence of the screening length on ion concentration and Bjerrum length is the opposite of that predicted by mean-field Debye-H\"{u}ckel theory. 

The data points in Figure \ref{Fig:ExperimentalFigureThree} differ in both ion diameter, concentration and Bjerrum length. To provide a clean test of the scaling relationship, Equation (\ref{scalinglaw_dim}), we perform a new set of SFB measurements where the Bjerrum length is varied by changing the solvent polarity thus the dielectric constant, but other variables are fixed by choosing the same salt and the same concentration. The SFB technique and detailed experimental procedures have been described elsewhere \cite{perkin2006forces,lee2017FD}. For reason of solubility, we used 2M solutions of 1-butyl-1-methylpyrrolidinium bis[(trifluoromethyl)sulfonyl]imide (abbreviated [C$_4$C$_1$Pyrr][NTf$_2$], Iolitec 99.5 \%) in propylene carbonate (Sigma-Aldrich, anhydrous 99.7 \%), dimethyl sulfoxide (Sigma-Aldrich, anhydrous 99.9\%), acetonitrile (Sigma Aldrich, anhydrous 99.8\%), benzonitrile (Sigma-Aldrich, anhydrous 99\%) and butyronitrile (Fluka, purity $\geq$99\%).  

\begin{figure}
\centering
\subfigure[]{\includegraphics[scale=0.8]{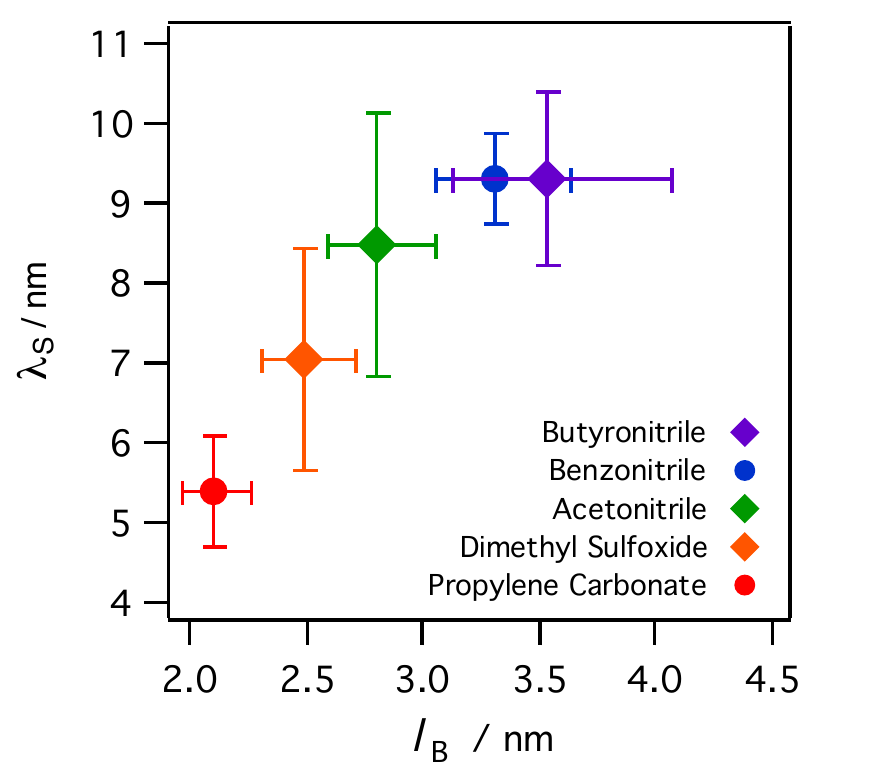}}
\subfigure[]{\includegraphics[scale=0.28]{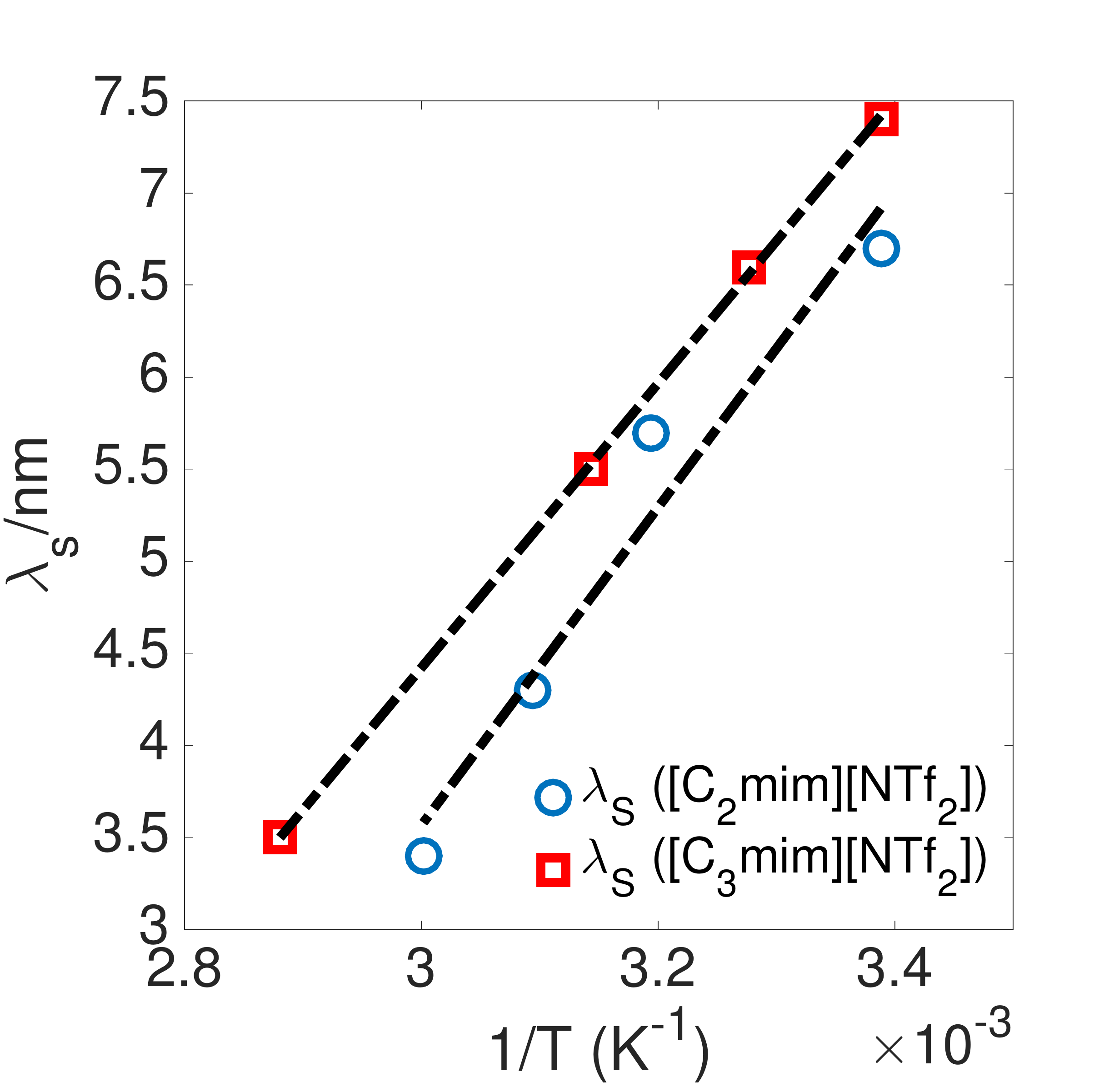}}
\caption{The measured screening length, $\lambda_S$, increases linearly with $l_B$: (a) Each data point corresponds to a 2M solution of [C$_4$C$_1$Pyrr] [NTf$_2$] in a different solvent and therefore different dielectric constant. Dielectric constants for the 2M solutions are calculated using effective medium theory \cite{Bergman1978}. The vertical error bars arise from scatter between the experimental decay length measured in different experiments and different force profiles in the same experiment. The horizontal error bar corresponds to scatter in the literature values of dielectric constant. See ref \cite{lee2017FD} for an expanded discussion. (b) $\lambda_S \sim 1/T$ for pure ionic liquids $\mathrm{[C_2 mim][NTf_2]}$ and $\mathrm{[C_3 mim][NTf_2]}$. The data (open circles and squares) are taken from ref \cite{gebbie2015long}. The dotted lines are the lines of best fit.}
\label{Fig:ExperimentalFigureFour}
\end{figure}

Figure \ref{Fig:ExperimentalFigureFour}(a) shows that the measured screening lengths are consistent with a linear increase with the Bjerrum length, in agreement with Equation (\ref{scalinglaw_dim}). The dielectric constant of the electrolyte solution is calculated using effective medium theory \cite{Bergman1978}. Another way to vary the Bjerrum length is by changing the temperature. Figure \ref{Fig:ExperimentalFigureFour}(b) reanalyses SFB data from  \cite{gebbie2015long} for pure ionic liquids $\mathrm{[C_2 mim][NTf_2]}$ and $\mathrm{[C_3 mim][NTf_2]}$. The screening length increases linearly with $1/T$, thus showing once again the linear relationship between the Bjrerrum length and screening length for concentrated electrolytes (assuming that the dielectric constant is independent of temperature). In the Supplemental Material we estimate the temperature-dependent dielectric constant and show that the scaling is robust.  

Thus far we showed that the screening length follows the scaling relationship Equation (\ref{scalinglaw_dim}) and is independent of the ion chemistry. A lingering question is whether the the screening length is a surface effect.  To address this question we turn to inspect the relation between screening length and activity coefficient, a bulk electrolyte quantity. If $\lambda_S$ is a bulk quantity, it is reasonable to posit, by analogy with the Debye--H\"{u}ckel theory, that the potential of mean force between ions, $v(r)$, decays exponentially with $\lambda_S$ as the decay length \cite{lee2017FD}, i.e.
\begin{equation}
v(r) = \begin{cases}
 - \frac{1}{\epsilon r} \frac{e^{a/\lambda_S}}{1+a/\lambda_S} e^{- r/\lambda_S} & a<r, \\
-\frac{1}{\epsilon a} \frac{1}{1+a/\lambda_S} & 0<r<a.
\end{cases}
\label{DH_solution}
\end{equation}
Assuming Equation (\ref{DH_solution}) holds, the electrostatic contribution to the excess chemical potential is 
\begin{equation}
\frac{\mu_{\mathrm{ex}}}{k_B T} = - \frac{1}{2} \frac{ l_B}{\lambda_S+ a}. 
\label{chemical_pot}
\end{equation} 
We note that the Yukawa form accounts for the spherical geometry of the ions, and Equation (\ref{DH_solution}) implies an exponentially decaying force between two planar surfaces with decay length $\lambda_S$, as observed in surface force measurements.  The prefactor in Equation (\ref{DH_solution}) is a simple consequence of charge neutrality \cite{lee2017FD}, and Equation (\ref{chemical_pot}) is the same as the Debye-H\"{u}ckel solution for the excess chemical potential except with the Debye length replaced by $\lambda_S$ \footnote{We refer the reader to ref \cite{lee2017FD} for an alternative justification based on relating the experimentally measured screening length to the free energy of charge fluctuations in a concentrated electrolyte.}. The excess chemical potential (or the activity coefficient $\gamma = e^{\mu_{\mathrm{ex}}/k_B T}$) of simple inorganic salts have been extensively tabulated in the classic electrochemistry literature \cite{robinson2002electrolyte} using bulk measurements. Therefore, comparing the excess chemical potential predicted by the measured screening length $\lambda_S$ and Equation (\ref{chemical_pot}), and the excess chemical potential that is independently measured, allows us to determine whether the measured screening length is a bulk quantity. 

\begin{figure}
\centering
\includegraphics[scale=0.35]{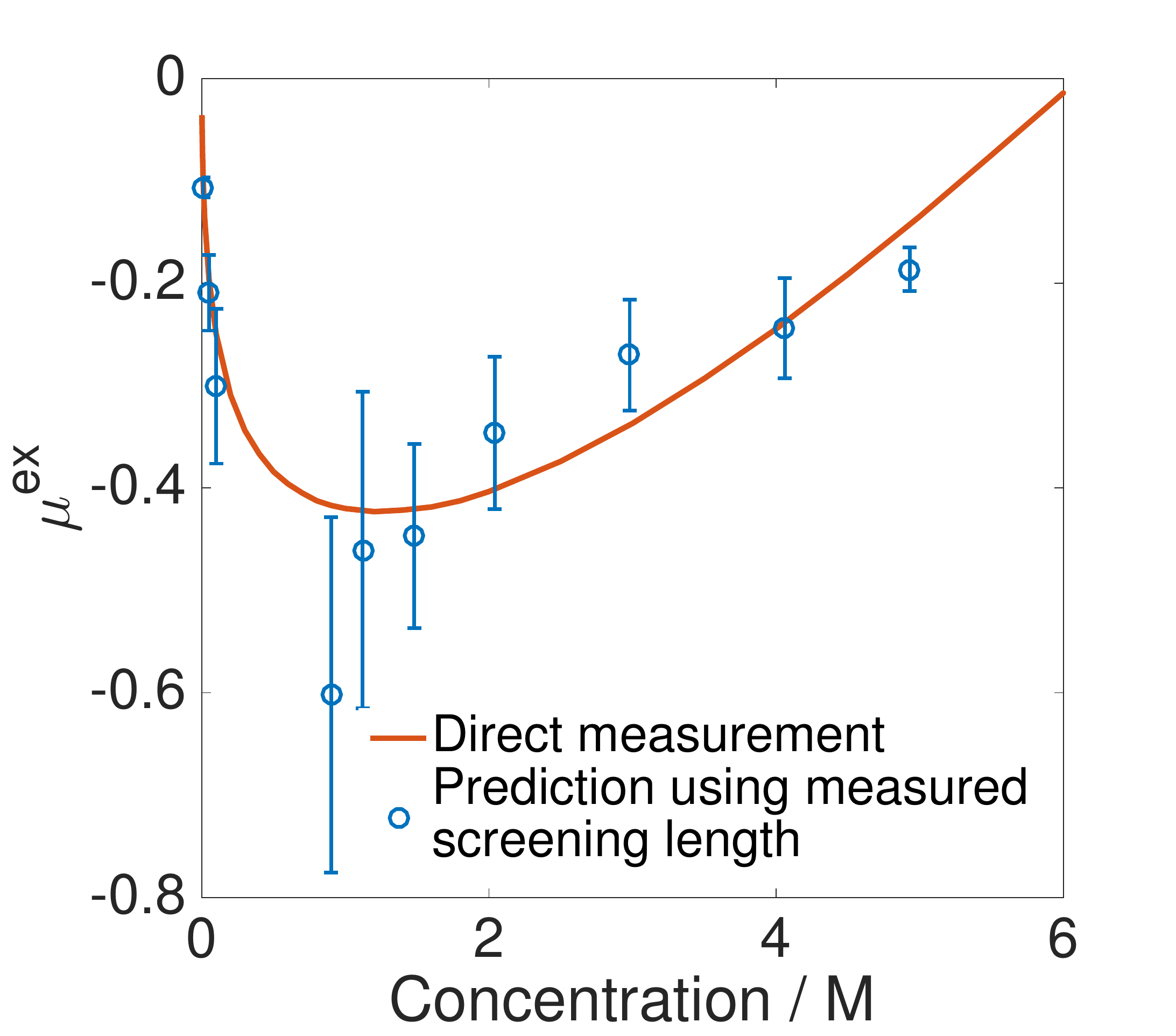} 
\caption{The excess chemical potential of aqueous sodium chloride solutions predicted using Equation (\ref{chemical_pot}) and the experimentally measured screening length agrees with direct measurements \cite{hamer1972osmotic}.}
\label{activity_coeff}
\end{figure}

Figure \ref{activity_coeff} shows that the measured screening length predicts the concentration dependence of the excess chemical potential of aqueous $\mathrm{NaCl}$, supporting that hypothesis that the measured screening length is indeed a bulk quantity. We focus on $\mathrm{NaCl}$ as screening length measurements in the literature for other inorganic salts are scarce. The upturn in excess chemical potential is commonly attributed to excluded-volume interactions \cite{robinson2002electrolyte} or the combined effects of ion-solvent interactions and ion-ion correlations \cite{vincze2010nonmonotonic,valisko2014effect,valisko2015unraveling,shilov2015role,valisko2015comment,valisko2017activity}. Here, we show that the upturn can be related to experimental measurements of the screening length as a function of concentration. Our model is conceptually consistent with previous theoretical approaches as the screening length reflects both ion-ion and ion-solvent correlations.

%Our model is conceptually consistent  \footnote{The major difference is that we take the \emph{experimentally measured} screening length as an input in order to test whether  the screening length is a bulk property. As such, we cannot decompose the chemical potential explicitly into ion-ion and ion-solvent terms because the screening length is a result of both interactions.} with previous theoretical approaches as the screening length reflects ion-ion correlation, which our scaling argument suggests is mediated by solvent as charge carriers, as shown in the discussions below. }

We finally turn to deriving a scaling theory for the screening length. We first consider a thought experiment: Suppose we put a grain of table salt, an ionic crystal, between two charged surfaces and ask whether the salt crystal screens the electric field. The answer is evidently \textit{no} because the ions are immobile and thus the crystal behaves as a dielectric slab. Now, suppose the crystal contains Schottky defects. Charge transport in such a defect-laden ionic crystal occurs via ions hopping onto defect sites. Alternatively, reminiscent of particle-hole symmetry, one could view the defect itself as the charge carrier \cite{kurosawa1957melting,hainovsky1995simple,zimmer2000charge}. Defects in the sub-lattice of the cations behave as negative charges, and defects in the sub-lattice of the anions behave as positive charges \cite{roberts1977surface}. The system would be able to screen an external electric field, but the charge carrier density that enters into the Debye length is the defect concentration rather than the ion concentration.

An ionic crystal is an extreme example of a correlated Coulomb melt where the ions are translationally immobile. A concentrated electrolyte behaves similarly to an ionic crystal in the sense that the electric potential felt by an ion due to all other ions is significantly greater than thermal fluctuations. The role of Schottky defects is played by the solvent molecules. Although solvents are charge-neutral molecules, they disrupt ion-ion correlation by freeing up a site that would have been occupied by an ion. Therefore, solvent molecules acquire an effective charge analogous to a defect in an ionic crystal. Indeed, the coupling between solvent concentration fluctuations and charge fluctuations is observed in molecular dynamics simulations of ionic liquid capacitors \cite{uralcan2016concentration}. 

We can put the physical intuition suggested above in a more quantitative footing by rewriting the Debye length in terms of ``defect'' concentration 
\begin{equation} 
\lambda_S =  (4 \pi \tilde{q}_{\mathrm{solv}}^2 l_B c_{\mathrm{solv}})^{-1/2}, 
\label{debyehuckel_defect}
\end{equation}
where $c_{\mathrm{solv}}$ is the concentration of solvent molecules, and $\tilde{q}_{\mathrm{solv}}^2$ is the mean-squared effective charge of a solvent molecule relative to the charge of an ion; the mean charge of a ``defect'' is zero in a symmetric electrolyte because it is as likely for a solvent molecule to be in the ``cation sub-lattice'' as in the ``anion sub-lattice''. Assuming the system is incompressible, $c_{\mathrm{solv}}  = c_{\mathrm{tot}} - c_{\mathrm{ion}}$, where $c_{\mathrm{tot}}$ is the total concentration of the system which is assumed to be independent of ion concentration. 

The next step is to estimate the effective mean-squared charge of a solvent molecule, or ``defect'', in this concentrated ionic system. Qualitatively, the defect takes the position of an ion in this correlated ionic system, and as such the energy of creating a defect must be comparable to the fluctuation energy of the ionic system per ion. The energy of a defect scales as $E_{\mathrm{defect}} \sim \tilde{q}_{\mathrm{solv}}^2$. This can be seen via symmetry (the defect energy is symmetric with respect to the charge of the defect), or by noting that a uniformly charged sphere of net charge $q$ has a self-energy that scales as $\sim q^2$. 

The energy density of the ion system can be derived using dimensional analysis: the only relevant electrostatic lengthscale in a system where Debye-H\"{u}ckel screening is negligible is the Bjerrum length. Therefore, one would expect the energy density $e_{\mathrm{ion}} \sim  l_{B}^{-3}$ from dimensional analysis. This estimate is analogous to the fluctuation energy for a dilute electrolyte which is known to scale as $\sim \lambda_D^{-3}$ \cite{safran1994statistical}, except the role of the Debye length in dilute electrolytes is replaced by the Bjerrum length in concentrated electrolytes because Debye screening is suppressed by strong ion-ion correlation. The electrostatic energy per ion is therefore  $E_{\mathrm{ion}} \sim a^3 e_{\mathrm{ele}} \sim (a/l_B)^3$. Equating  $E_{\mathrm{ion}}$ with $E_{\mathrm{defect}}$ gives the scaling relationship  
\begin{equation}
\tilde{q}_{\mathrm{solv}}^2 \sim \left(\frac{a}{l_B} \right)^3. 
\label{scale-charge}
\end{equation}
This charge scaling shows the important physics that strong ionic correlations (large Bjerrum length) suppresses thermal fluctuations in the system, and therefore the mean-squared charge of a defect which is acquired through fluctuations.   

Substituting (\ref{scale-charge}) and the incompressibility constraint into Equation (\ref{debyehuckel_defect}), we obtain 
%\begin{align}
%\Lambda_D &\sim  (4 \pi  ( c_{\mathrm{tot}} - c_{\mathrm{ion}})a^3/l_B^2)^{-1/2} \nonumber \\ 
%&\approx  (4 \pi c_{\mathrm{tot}}a^3/l_B^2)^{-1/2} + \frac{1}{2 \sqrt{4 \pi} (c_{\mathrm{tot}} a^3)^{3/2} } l_B c_{\mathrm{ion}}a^3 , 
%\label{mod_DH}
%\end{align}  
\begin{align}
\frac{\lambda_S}{\lambda_D} &\sim  \frac{(4 \pi  ( c_{\mathrm{tot}} - c_{\mathrm{ion}})a^3/l_B^2)^{-1/2}}{(4 \pi  c_{\mathrm{ion}} l_B)^{-1/2}} \nonumber \\ 
&= C \left( \frac{a}{\lambda_D} \right)^3, 
\label{mod_DH}
\end{align}  
where $C =(\phi_{\mathrm{ion}} \sqrt{\phi_{\mathrm{tot}}-\phi_{\mathrm{ion}}})^{-1}$ and $\phi_{\mathrm{ion}/\mathrm{tot}} = c_{\mathrm{ion}/\mathrm{tot}} a^3$. Noting that the prefactor $ C$ is approximately constant for concentrated electrolytes $2 c_{\mathrm{tot}}/3 \ll c_{\mathrm{ion}} \ll c_{\mathrm{tot}}$, Equation (\ref{mod_DH}) shows that the scaling obtained from this simple physical picture agrees with the empirically observed scaling relationship, Equation (\ref{scalinglaw}). For ionic liquids, although there is no solvent molecule per se, the internal degrees of freedom in the ions, in particular the alkyl chains on the cation, could perform the role of the solvent by disrupting order in the strongly correlated ionic melt. 

We next consider the ion concentration at which this ``ionic crystal'' analogy becomes appropriate. The discussion above suggests that the ionic crystal regime is reached when the typical ion-ion electrostatic interaction energy is greater than $k_B T$. We can put this intuition in a more quantitative footing: Consider a spherical blob of electrolyte of radius $R$ in the bulk electrolyte. Modelling the blob as an uniformly charged sphere, the fluctuation energy of the blob is given by 
\begin{equation} 
E_{\mathrm{fluct}} \sim k_B T l_B \frac{\left<Q^2\right>}{R}
\label{fluct_energy}
\end{equation} 
where $Q$ is the charge of the blob. If charge fluctuations in the blob follows Gaussian statistics, then $\left<Q^2\right> \sim N_{\mathrm{ion}}$ where $N_{\mathrm{ion}}$ is the number of ions in the blob, which in turn is related to the bulk density via $N_{\mathrm{ion}} \sim c_{\mathrm{ion}} R^3$. Therefore $E_{\mathrm{fluct}} \sim k_B T l_B c_{\mathrm{ion}} R^2 $ and the fluctuation energy increases with the blob size. The minimal blob size is obviously the ion diameter, and the strong correlation regime is reached when the fluctuation energy of even this minimal blob is above $k_B T$. In other words 
\begin{equation}
 l_B c_{\mathrm{ion}} a^2 \sim 1
 \label{fluct_energy_kt}  
\end{equation} 
Equation (\ref{fluct_energy_kt}) can be rewritten as $a/\lambda_D \sim 1$, agreeing with the experimental results (c.f. Figure \ref{Fig:ExperimentalFigureThree}). 

In summary, we presented a series of experimental results showing that the interaction between charged surfaces in a concentrated electrolyte decays exponentially with a decay length that follows the scaling relationship $\lambda_S \sim l_B c_\mathrm{ion} a^3$, where $l_B$ is the Bjerrum length, $c_\mathrm{ion}$ the ion concentration and $a$ the ion diameter. This scaling relationship is robust to varying the chemical functionalities or molecular features of the ions, and holds for both ionic liquid solutions and alkali halide solutions. We show that the screening length is a bulk property of the electrolyte by relating the screening length to the independently-measured excess chemical potential of electrolyte solutions. The scaling relationship between the screening and ion concentration and Bjerrum length supports a novel physical picture: in a concentrated electrolyte, ions are strongly correlated and it is the neutral solvent molecules that acts as charge carriers; the solvent molecules acquire an effective charge through thermal fluctuations. We show that the empirically observed scaling relationship $\lambda_S \sim l_B c_\mathrm{ion} a^3$ follows naturally from this physical picture.

\acknowledgments
The authors are very grateful to S Safran for many insightful discussions. The authors would like to thank the participants of the ``Chemical Physics of Electroactive Materials: Faraday Discussions'' meeting in Cambridge, UK ($\mathrm{10^{th} - 12^{th}}$ April, 2017). We have presented a preliminary version of the arguments in this Letter in the meeting \cite{lee2017FD}; our conference paper, together with comments and questions we received from our colleagues, will be published in the conference proceedings in due course. AMS is supported by a Doctoral Prize from the EPSRC. AAL is supported by a UK-US Fulbright Fellowship to Harvard University and the George F. Carrier Fellowship. SP and CPM are supported by The Leverhulme Trust (RPG-2015-328) and the ERC (under Starting Grant LIQUISWITCH). 

\bibliography{underscreening_refs} 
\end{document}